\documentclass[a4paper,11pt]{article}

\usepackage{jcappub} 

\usepackage[T1]{fontenc} 

\usepackage{subfigure}

\newcommand{\be}{\begin{equation}}
\newcommand{\ee}{\end{equation}}

\title{\boldmath Turnaround radius in an accelerated universe with quasi-local mass}

\author[a,1]{Valerio Faraoni,\note{Corresponding author.}}
\author[b]{Marianne Lapierre-L\'eonard}
\author[a]{and Angus Prain.}

\affiliation[a]{Physics Department and STAR Research Cluster, Bishop's University, Sherbrooke, Qu\'ebec, Canada J1M 1Z7}
\affiliation[b]{Physics Department, Bishop's University, Sherbrooke, Qu\'ebec, Canada J1M 1Z7}

\emailAdd{vfaraoni@ubishops.ca}
\emailAdd{mlapierre12@ubishops.ca}
\emailAdd{angusprain@gmail.com}

\abstract{We apply the Hawking-Hayward quasi-local energy construct to obtain in a rigorous way the turnaround radius of cosmic structures in General Relativity. A splitting of this quasi-local mass into local and cosmological parts describes the interplay between local attraction and cosmological expansion.}


\begin{document}
\maketitle
\flushbottom

\section{Introduction}

The accelerated expansion of the universe has led cosmologists
to postulate the existence of a mysterious dark energy
as an explanation, a fluid with exotic properties (very negative 
pressure $P \simeq - \rho$, where $\rho$ is the energy density) \cite{AmendolaTsujikawa}. Alternatives 
to this {\it ad hoc} dark energy have been proposed, including 
the backreaction of perturbations on the cosmic dynamics \cite{Buchert, Buchert2, Rasanen}, 
living in a giant void \cite{Boleiko}, or 
modifying altogether gravity at large scales \cite{Review1, Review2, Review3}. 
Even if one accepts dark energy as an explanation,
there are a plethora of models and it is important
to design observational tests which can discriminate
between dark energy and modified gravity, or 
between various dark energy models. 

Recently, 
the concept of turnaround radius has been proposed 
as an interesting and promising way to test dark 
energy. Since an accelerated cosmic expansion 
opposes the collapse of local overdensities and the 
formation of cosmological structures, it is not surprising 
that, in an accelerating Friedmann-Lema\^itre-Robertson-Walker 
(FLRW) universe with a \emph{spherical} inhomogeneity, 
there is an upper bound to the radius of an 
overdensity of a certain mass, beyond which 
this matter cannot collapse but can only expand.  
Assuming, for reference, that the universe in which 
we live is approximately described by the standard
$\Lambda$-Cold Dark Matter ($\Lambda$CDM) scenario of General
Relativity with a cosmological constant or dark energy, it 
has been suggested that the $\Lambda$CDM scenario can be 
tested by studying whether the mass-radius relation
of structures in the cosmos respects a theoretical
prescription \cite{Souriau, Stuchlik1, Stuchlik2, Stuchlik3, Stuchlik4, Mizony05, Stuchlik5, Roupasetal, Nolan2014, PT, PTT}. 

\begin{figure}
\centering
\subfigure[]{
\includegraphics[scale=0.37]{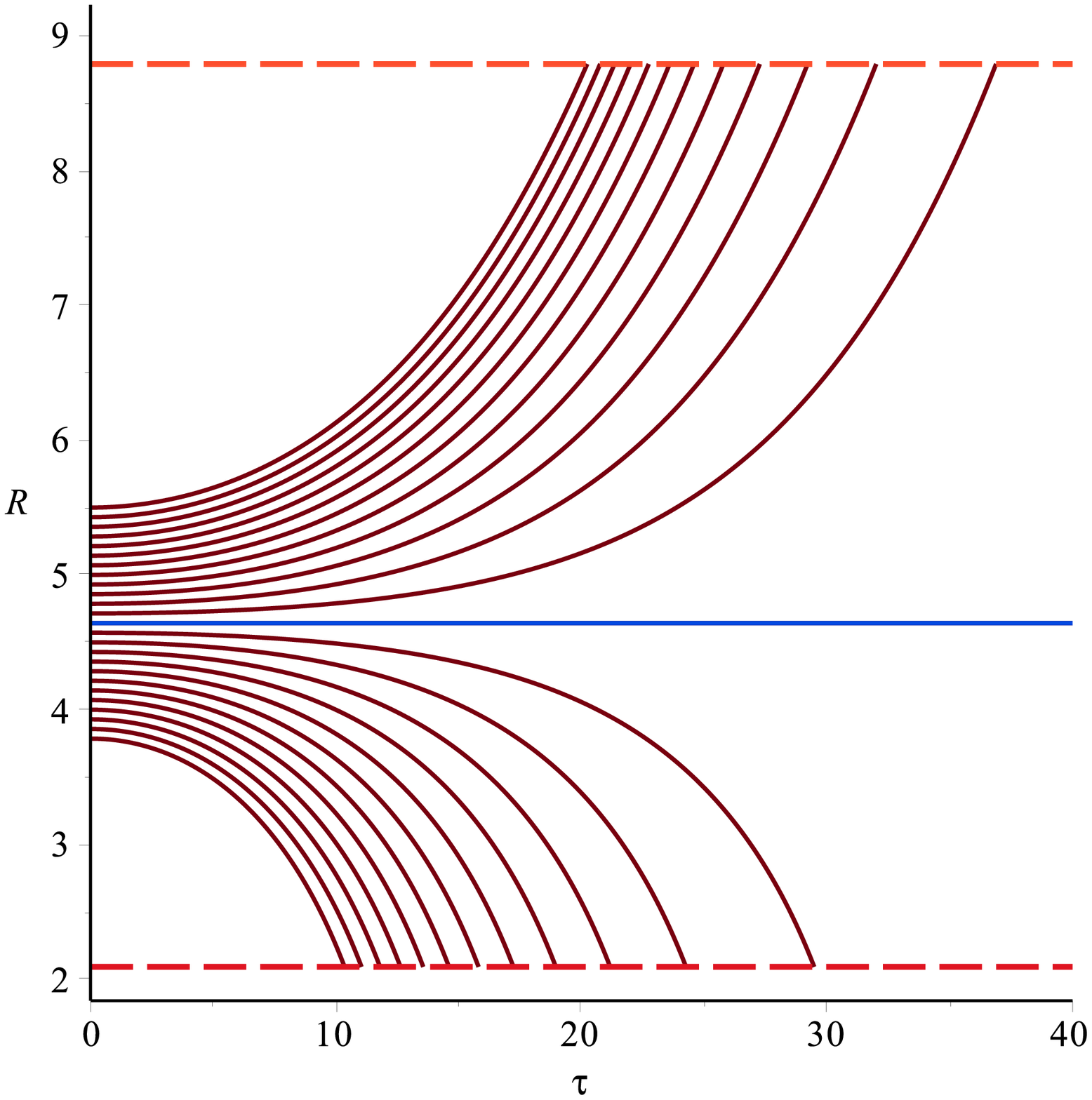}
\label{sF:zero}}
\subfigure[]{
\includegraphics[scale=0.37]{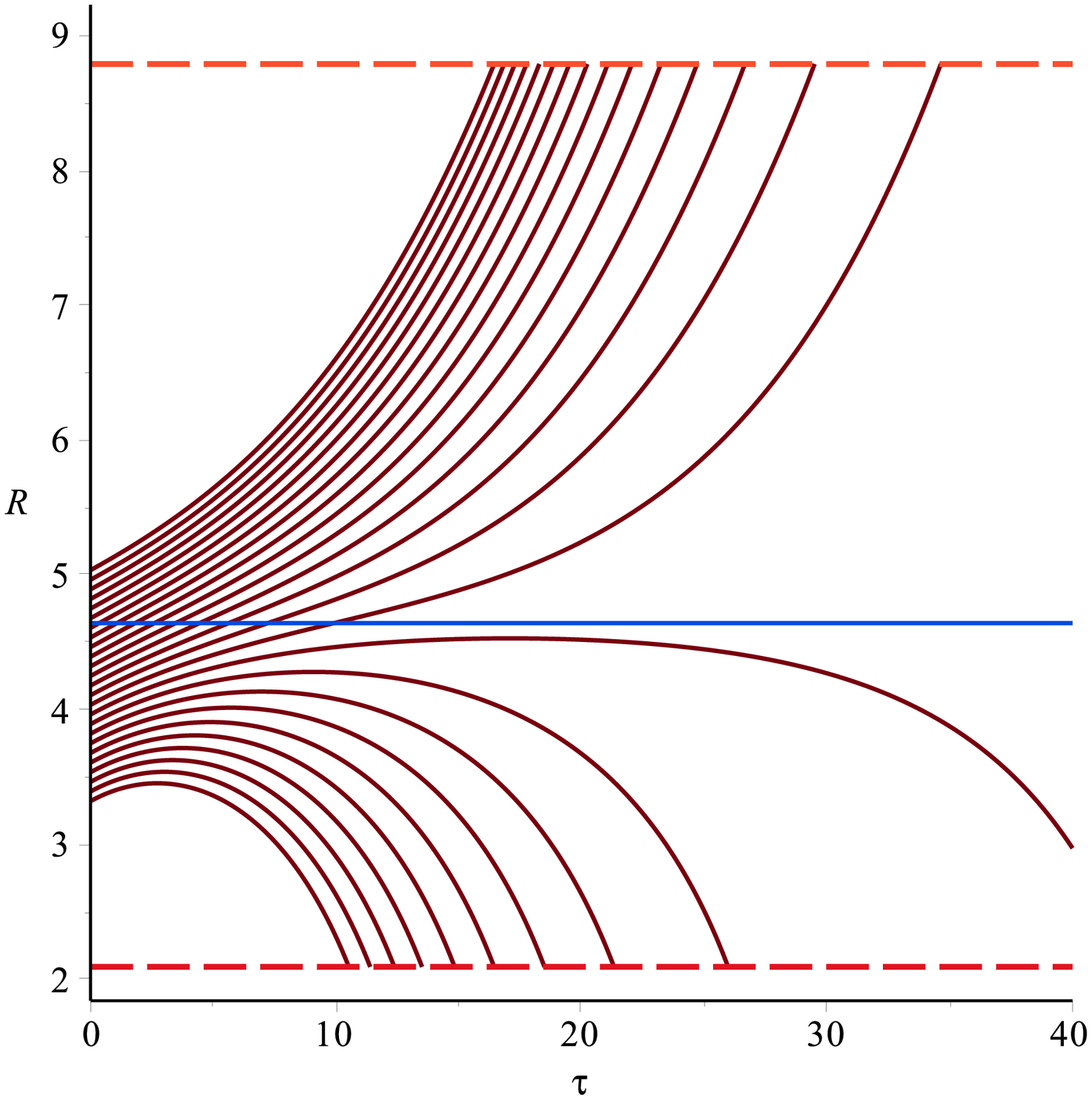}
\label{sF:out} }
\subfigure[]{
\includegraphics[scale=0.37]{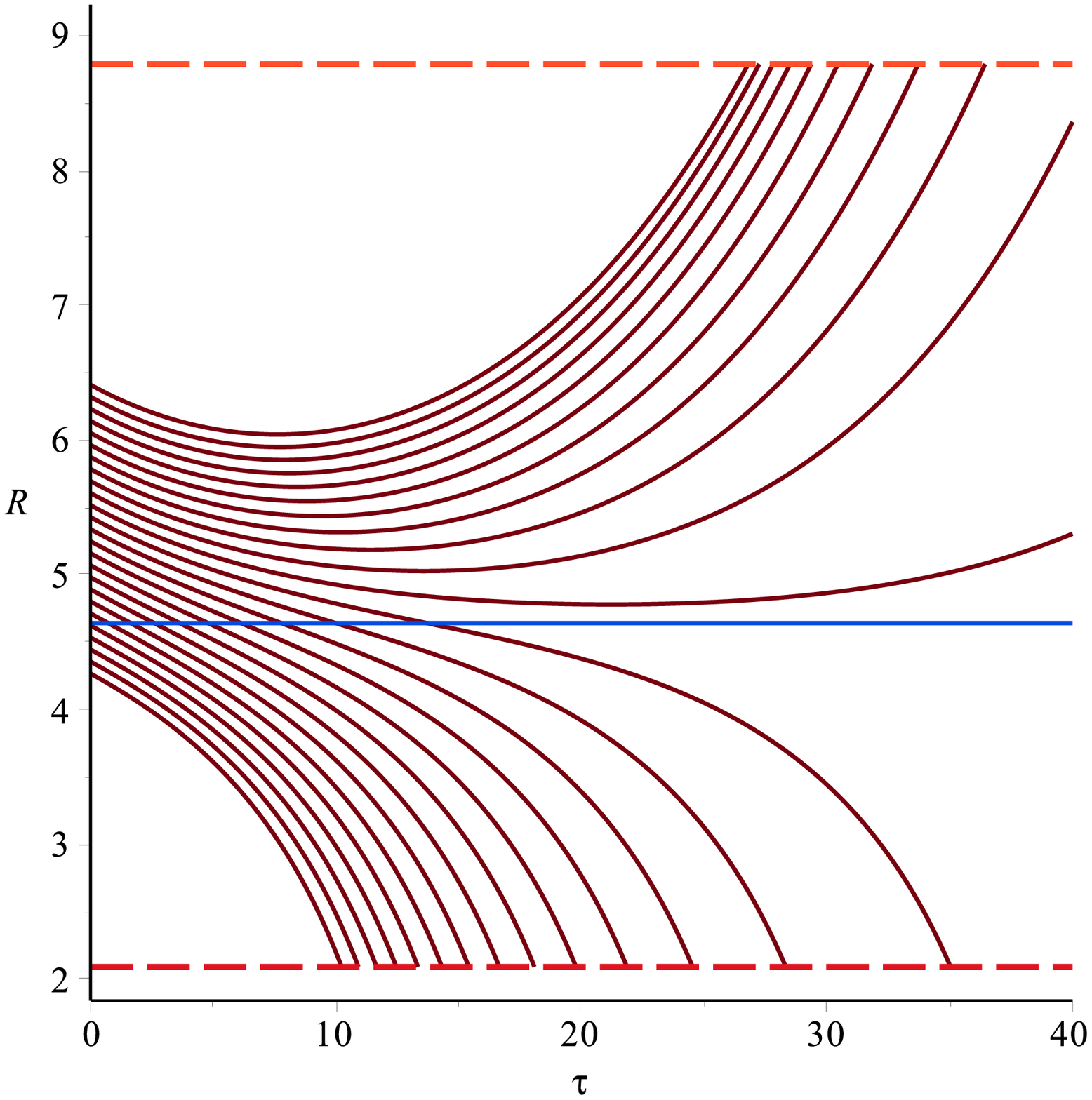}
\label{sF:in}}
\caption{Radial trajectories of massive test particles in the SdS spacetime subject to zero \ref{sF:zero}, outgoing \ref{sF:out} and in-falling \ref{sF:in} initial velocities. The solid blue line is the location of the critical radius $R_c$ \eqref{turnaround-lambda}.   To make these figures we used the illustrative parameter values $M=1$ and $H=0.1$ for which both the cosmological and black hole horizons are present (dashed lines). $\tau$ is the proper time (an affine parameter) along the geodesics.  \label{F:SdS}}
\end{figure}

To be more precise, consider
a spherical configuration of mass in an accelerating FLRW
universe: when the outer layers of this material configuration
 reach zero radial acceleration and collapse under the self-gravity of the perturbation,
it is said that the perturbation has reached its \emph{turnaround
radius} (\cite{PT, PTT}, see also \cite{BlauRollier, DG1, DG2, DG3, DG4, DG5, DG6}). In a decelerated universe there is 
no upper bound on the turnaround radius, but in a 
de Sitter background with constant Hubble parameter $H=\sqrt{\Lambda/3}$
(where $\Lambda > 0$ is the cosmological constant), there exists the upper bound 
\be \label{turnaround-lambda}
R_c = {\left( \frac{3GM}{\Lambda} \right)}^{1/3}
\ee
on the turnaround radius, where $R$ is the 
areal radius of the Schwarschild-de Sitter (SdS) line element,
\be
ds^2 = -\left(1- \frac{2M}{R} - H^2 R^2 \right) dt^2 + \frac{dR^2}{1- \frac{2M}{R} - H^2 R^2} + R^2 d\Omega^2 _{(2)}
\ee
in static coordinates, $d\Omega^2 _{(2)} = d\vartheta ^2 + \sin^2 \vartheta d\varphi^2$ is the 
line element on the unit 2-sphere, and $H=\sqrt{\Lambda/3}$ \cite{PT, PTT}.  In Fig.\ref{F:SdS} we plot the trajectories of test particles placed at various radii in the SdS spacetime, with initially zero \ref{sF:zero}, outgoing \ref{sF:out} and in-falling \ref{sF:in} initial velocities. 
The figure shows that the critical radius is not an absolute boundary in the same way a horizon is since timelike geodesics can cross it in either direction. The critical radius is that point beyond which, if you cross outside of it in geodesic motion, you will never cross back without proper acceleration, and \textit{vice-versa}.  This one-way property is also directly visible from the geodesic equation for radial timelike trajectories, which can be shown to be given by
\be
\ddot{R}(\tau)=\left(R^3-R_c^3\right)\frac{H^2}{R^2} \label{E:SdS_geodesic},
\ee
where an overdot denotes differentiation with respect to the proper time $\tau$ parametrizing timelike geodesics. We see that for $R>R_c$ we have $\ddot{R}>0$ and for $R<R_c$ we have $\ddot{R}<0$. This is the defining property of the turnaround radius.\footnote{The other geodesic equation for radial motion is 
\be
\ddot{t}(\tau)-2\left(R^3-R_c^3\right)\frac{H^2}{R^2}\left(1-\frac{2M_\text{MSH}}{R}\right)^{-1}\dot{t}(\tau)\dot{R}(\tau)=0 , 
\ee
where $M_\text{MSH}$ is the quasi-local mass of a sphere of radius $R$.
}

The concept of turnaround radius can be generalized 
to arbitrary (but accelerating) FLRW universes \cite{PTT}.
In Ref.~\cite{PTT}, also a Lema\^itre-Tolman-Bondi metric
\be
ds^2 = - dt^2 +\frac{R'^2(t,r)}{1+f(r)} dr^2 +R^2(t,r) d\Omega^2 _{(2)}
\ee
is studied, where $f(r)$ is an arbitrary function
related to an initial density profile, and a cosmological
constant $\Lambda$ is present in addition to the usual dust fluid (a prime denotes differentiation with respect to $r$). The authors study shells of
areal radius $R$ and derive their radial acceleration 
\be
\ddot{R} = - \frac{G \mathcal{M}(r)}{R^2} + \frac{\Lambda R}{3} \,,
\ee
obtaining a turnaround radius
\be
R_c (t_c,r_c) = {\left( \frac{3G \mathcal{M}(r_c)}{\Lambda} \right)}^{1/3} \,.
\ee
Here $\mathcal{M} (r) = 4 \pi \int ^R _0 R^2 \rho dR$ is the well-known Lema\^itre 
mass and $\rho$ is the density of the cosmic dust in the Lema\^itre-Tolman-Bondi spacetime.

\begin{figure}
\centering
\subfigure[]{\label{sf:a}
\includegraphics[scale=0.35]{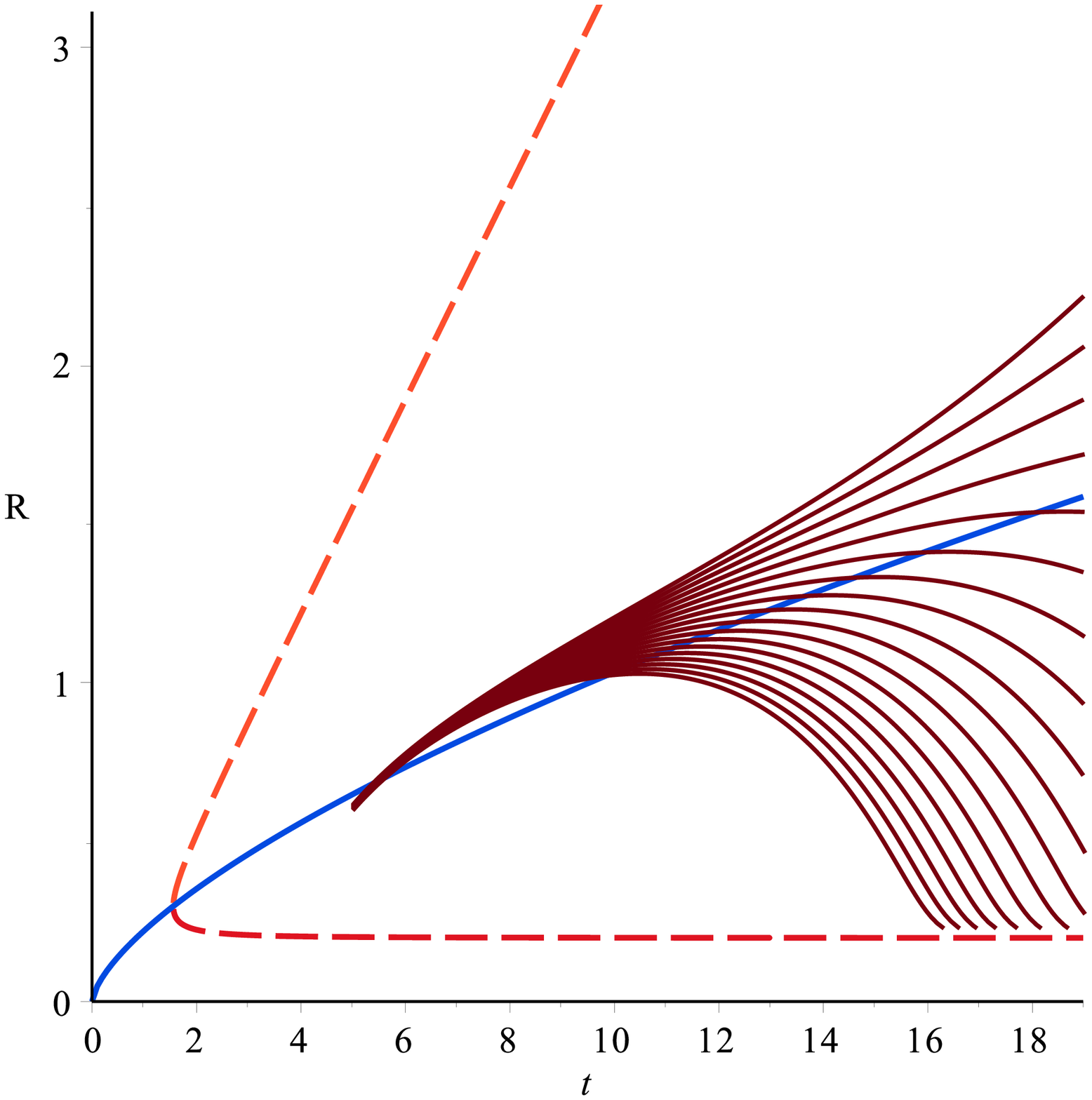}
}
\subfigure[]{\label{sf:b}
\includegraphics[scale=0.35]{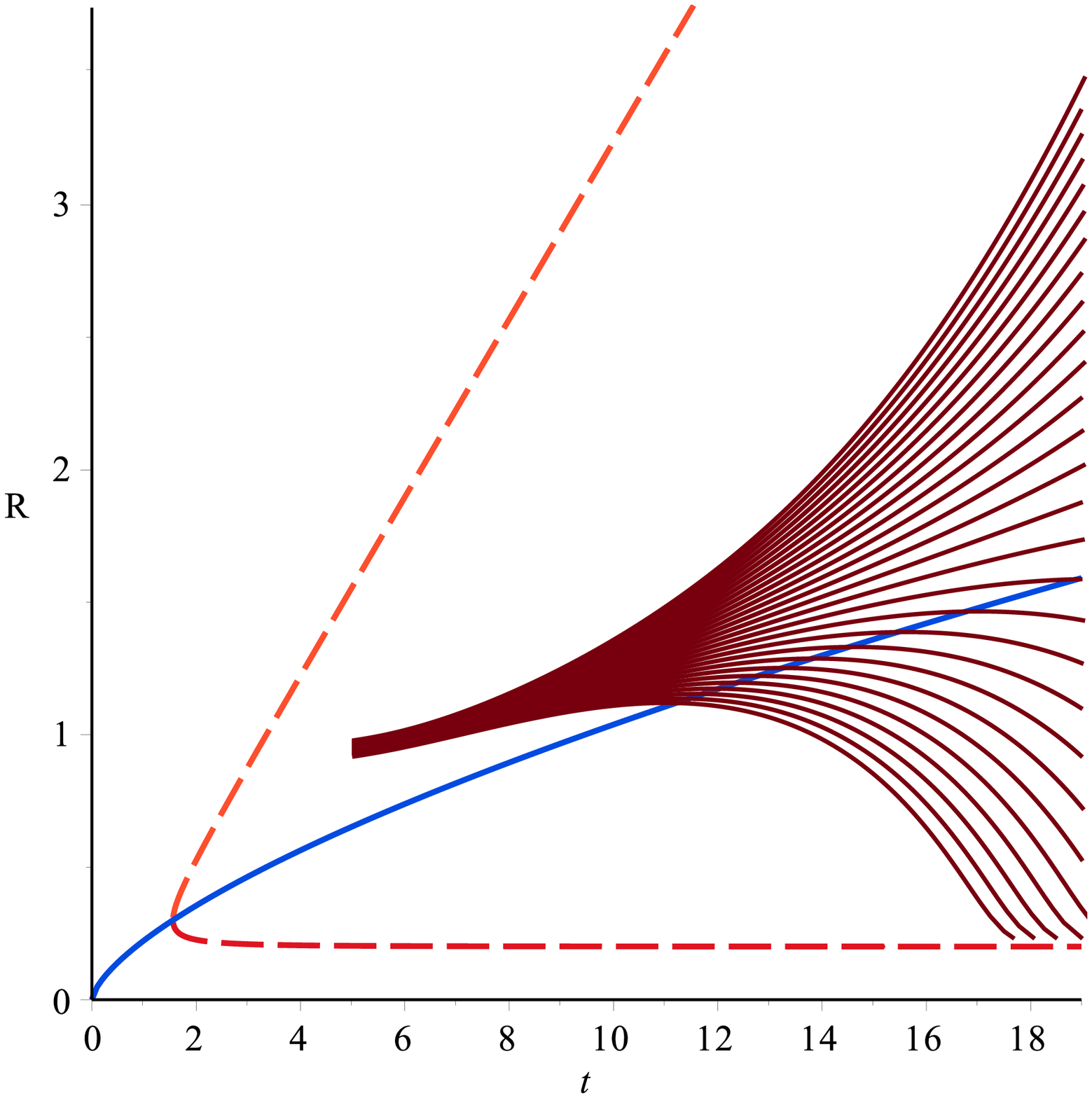}
}
\caption{Timelike geodesics with large positive (subfigure \ref{sf:a})  and small positive (subfigure \ref{sf:b}) initial velocities.  The critical radius is the solid blue line and the apparent horizons for McVittie spacetime with $a(t)=t^3$ are the dashed red and orange curves. t is the coordinate time.  \label{F:McVitte_geodesics}}
\end{figure}

The Lema\^itre-Tolman-Bondi model is only one of the 
possible choices to describe a spherical inhomogeneity
embedded in a FLRW universe: other choices exist, for
example the McVittie \cite{FaraoniJacques} and generalized McVittie \cite{McVittie}
metrics, which have been the subject of recent attention \cite{Kleban, Lake1, Lake2, 
Roshina1, Roshina2, AndresRoshina, DSFG2012, DaSilvaGuarientoMolina2015, NiayeshDaniel, Horndeski, FaraoniGaoetc} and which describe a generalization of Schwarzschild-de Sitter spacetime to include a time-dependent Hubble parameter $H(t)$. The McVittie spacetime is described by the metric
\be
ds^2=-\left(1-\frac{2m}{R}-H^2R^2\right)dt^2+\left(1-\frac{2m}{R}\right)^{-1}dR^2- \frac{2HR}{\sqrt{1-\frac{2m}{R}}} \, dtdR+R^2d\Omega^2_{(2)}
\ee
with $H(t) \equiv \dot{a}(t)/a(t)$  the time-dependent Hubble parameter. The radial geodesic equation in McVittie spacetime is a generalization of eq.~\eqref{E:SdS_geodesic} and it can be shown to be
\be
\ddot{R}(\tau)=\left(R^3-R_c^3\right)\frac{H^2}{R^2}+\dot{H}R\sqrt{1-\frac{2m}{R}}\,\dot{t}^2(\tau)
\ee
Note that in the de Sitter case (when $a(t)=\exp(\mathcal{H}t)$ with $\mathcal{H}$ constant), the extra term proportional to $\dot{t}^2$ vanishes exactly.  Therefore, the turnaround property at $R=R_c$ is not an exact result in the McVittie case and the details depend on the precise evolution of the background as well as the time in which the sprinkling takes place. For power-law expansion $a(t)=t^p$ at the radius $R=R_c$ we have 
\be
\ddot{R}(\tau,R_c)\propto \frac{-1}{t^{8/3}(\tau)}\sqrt{1-\frac{2m}{R_c}}\,\dot{t}^2(\tau) \label{E:stricly},
\ee
which goes to zero at late times, establishing that the sphere $R=R_c$ is critical at late times. In Fig.~\ref{F:McVitte_geodesics} we show congruences of timelike geodesic trajectories starting near the time-dependent radius $R=R_c$  and the (time-dependent) apparent horizons to illustrate this discussion for McVittie. 
The figure shows that it is possible to find geodesics which escape but are subsequently captured by the critical sphere. Note that the result \eqref{E:stricly} shows that the exact critical radius is strictly larger than $R_c$ since $\ddot{R}(t,R_c)$ is always negative inside the horizons.

In any case, when one decides to describe 
a large-scale structure (assumed
to be spherical for simplicity) in a FLRW universe, the deviations of
the metric from an exact FLRW one are small (although the density contrast, {\em i.e.},
the ratio between the density of the local perturbation and the average
cosmological density can be large).
By adopting the conformal Newtonian gauge, the
perturbed FLRW metric is written as
\be
ds^2 = a^2(\eta)\left[ - (1+2\phi)d\eta^2 +(1-2\phi)\left( dr^2 +r^2 d\Omega^2 _{(2)} \right) \right] \,,
\ee
where $\eta$ is the conformal time of the FLRW background 
(defined by $dt=ad\eta$ in terms of the comoving time $t$),
$a$ is the scale factor, and $\phi$ is a post-Friedmannian 
potential describing the local inhomogeneity. Here we
neglect vector and tensor perturbations, as is customary 
to first order, and we restrict our  discussion to
linear order in the metric perturbations, which 
are assumed to be small everywhere. As usual,
the spatial derivatives of $\phi$ dominate over its time derivative,
and we confine ourselves to General Relativity (otherwise
two distinct potentials $\psi$ and $\phi$ would appear in the
metric coefficients $g_{00}$ and $g_{11}$).

In spherical symmetry $\phi = \phi(r)$ and it is possible to
introduce the areal radius
\be
R(\eta,r)=a(\eta) r \sqrt{1-2\phi(r)} \,.
\ee
The authors of \cite{PTT} derive the radial acceleration
of a spherical shell in comoving time
\be \label{radial_acceleration}
\ddot{R} = - \frac{4 \pi}{3} (\rho_E +3P_E)R - \frac{G \mathcal{M}(r)}{R^2} = \frac{\ddot{a}}{a} R - \frac{G \mathcal{M}(r)}{R^2} \,,
\ee
where an overdot denotes differentiation with respect to the comoving time of the FLRW 
background, $\rho_E$ and $P_E$ are the energy density and pressure of the
dark energy propelling the cosmic acceleration, respectively, 
\be
\mathcal{M}(r) = 4 \pi \int ^r _0 r'^2 \rho(r') dr'
\ee
is the mass contained within the shell of radius $R(t,r)$,
and the energy density $\rho$ includes both the homogeneous cosmological density
$\rho_E$ and the inhomogeneous density of the perturbation.\footnote{However, this expression for the mass is not written explicitly in \cite{PTT}.} It is then straightforward to
derive the radius $R_c$ corresponding to the situation
in which the two contributions, cosmological and
local, cancel out in the right hand side of eq.~(\ref{radial_acceleration}) \cite{PTT},
\be
R_c = {\left( - \frac{3 \mathcal{M}}{4 (1+3w) \pi \rho_E} \right)}^{1/3} \,,
\ee
where $w \equiv \frac{P_E}{ \rho_E} <-\frac{1}{3}$ is the equation of state parameter of 
dark energy. This expression is taken as the 
turnaround radius in a dark energy-dominated 
FLRW universe in which $H=H(t)$. For $w=-1$ (the value corresponding to a de Sitter background),
it reduces 
to the expression~(\ref{turnaround-lambda}). A similar expression defines the ``sphere of
influence'' used for the same purpose in the literature ({\em e.g.}, \cite{Bushaetal}).

Writing the perturbed FLRW metric in terms of the areal radius and the Misner-Sharp Hernandez quasi-local mass\footnote{We will introduce and motivate the quasi-local mass  more fully in section \ref{S:qlm}. Essentially, $M_\text{MSH}$ measures the total mass/energy inside a sphere of areal radius $R$, including gravitational contributions.}
\be
M_\text{MSH}=-\phi R+\frac{H^2 R^3}{2}-H^2 R^3 \phi
\ee
we have
\begin{align}
ds^2&=-\left(1-\frac{2M_\text{MSH}}{R}\right)dt^2+\left(1-\frac{2M_\text{MSH}}{R}+H^2R^2\right)^{-1}dR^2+R^2d\Omega_{(2)}^2\notag \\
&\hspace{25mm}-\frac{2HR}{1-\frac{2M_\text{MSH}}{R}+H^2R^2}\,dR dt
\end{align}
which is ``McVittie-like'', since the McVittie metric is written in terms of the quasi-local mass as
\begin{align}
ds^2&=-\left(1-\frac{2M_\text{MSH}}{R}\right)dt^2+\left(1-\frac{2M_\text{MSH}}{R}+H^2R^2\right)^{-1}dR^2+R^2d\Omega_{(2)}^2 \notag \\
&\hspace{25mm}-\frac{2HR}{\sqrt{1-\frac{2M_\text{MSH}}{R}+H^2R^2}}\,dR dt\, .
\end{align}
We see that the only difference is in the off-diagonal terms in the metric which could arguably be equal to the level of approximation we employ in the FLRW metric.\footnote{Since $M_\text{MSH}=\mathcal{O}\left(\phi\right)$ we have $H^2R^2=\mathcal{O}\left(\phi\right)$ and hence the numerator is $\mathcal{O}\left(\phi^{1/2}\right)$ and the denominator can be split into two products of the square root of itself, one of which contributes $1+\mathcal{O}\left(\phi\right)$ to the numerator and is therefore discarded.} As a linearized solution, however, the perturbed FLRW metric in ``McVittie form" should only be used far from the inner apparent horizon when it exists. Nevertheless, the geodesic structure near the critical radius and its approximate nature carry over from the McVittie case.

We note that:
\begin{enumerate}
\item There is a vast literature on the interplay between cosmological 
expansion and local physics (see \cite{CarreraGiulini} for 
a review), originating in the ``pure'' relativity community,
which seems to be mostly neglected by the more astrophysically-oriented community working on the turnaround radius
and studying real celestial objects. Here we attempt to bridge between these two
approaches and communities, in view of the fact
that the turnaround radius holds much promise for
testing the $\Lambda$CDM model. 

\item Discussions relying on a particular gauge (such as the conformal Newtonian gauge) are gauge-dependent
and one would like, instead, to have gauge-invariant characterizations.

\item The ``mass contained within a sphere'' is defined
in a rather hand-waving way in the literature on the 
turnaround radius: should this ``mass'' include
that of the cosmological fluid but not its pressure (as in
\cite{PTT})? Should it include the negative pressure of
dark energy? If not, why not? Should it include only
the local density ({\em i.e.}, energy density minus the cosmological density)? These
questions are not answered, nor asked explicitly, in the
literature on the turnaround test of $\Lambda$CDM. 
It is clear, however, that the turnaround spheres considered
contain significant amounts of dark energy and
a relativistic definition of gravitating mass-energy
in them should be used. In other words, the universe
of cosmology is relativistic and regions sufficiently large 
should {\em a priori} be treated with relativity and
not with Newtonian physics.
Indeed, the questions above are not trivial: researchers in General Relativity
have wondered for decades about the correct definition
of gravitating mass-energy for a given spacetime,
especially when the latter is not asymptotically
flat (as is the case in cosmology), and have come up
with several quasi-local approaches and mass
definitions (see \cite{Szabados} for a review). There is
now a large consensus that the ``correct'' notion is
the Hawking-Hayward quasi-local mass \cite{Hawking, Hayward}.
We have recently applied \cite{VMA} 
the Hawking-Hayward quasi-local mass $M_\text{HH}$ construct to
cosmological perturbations, giving a covariant splitting of this quantity into ``local'' and
``cosmological'' parts which can be used
to separate the gravitational effects of local perturbations
and cosmological background. Although Ref.~\cite{VMA}
studied the growth of structures in the dust-dominated era,
the formalism presented in it is general and can be immediately adapted to the
accelerated era, which is what we set out to do here.
Ref.~\cite{VMA} is not restricted to
spherical symmetry, although it includes the spherically
symmetric case as an example.

\end{enumerate}

Our goals include clarifying the issue of mass within 
a turnaround sphere and giving firmer foundations
to the concept of turnaround radius. Although, in practice,
there is at least a 10~--~30~$\%$ uncertainty in the determination
of masses and radii of the astronomical objects used
in the turnaround test of $\Lambda$CDM, vagueness
and uncertainty in the theory do not help
getting firm outcomes for the test.

It turns out that, in a FLRW universe dominated by 
dark energy with equation of state parameter $w \simeq -1$,
the difference between the more rigorously defined
turnaround radius and the one used in the astronomical
literature is minimal. The ambiguities inherent in 
the definition of ``mass'' are clarified (although the
astronomical uncertainties on the \emph{values} of these
masses, of course, persist). The ``mass''
is here defined in a completely gauge- and coordinate-independent 
way (independent of the conformal Newtonian
gauge chosen at the outset for the perturbed FLRW metric).

\section{A rigorous definition of turnaround radius \label{S:qlm}}
Here we approach the issue of cosmological accelerated
expansion versus local physics and that of a rigorous
definition of turnaround radius using the Hawking-Hayward 
quasi-local mass. Our assumptions are:

\begin{enumerate}
\item General Relativity is valid (otherwise the Hawking-Hayward 
quasi-local energy is not defined and the perturbed FLRW metric is not 
given by eq.~(\ref{CNG})).

\item We restrict ourselves to first order in the metric
perturbations (however, density fluctuations can be
large). The perturbed FLRW metric is written
in the Newtonian conformal gauge as
\be \label{CNG}
ds^2 = a^2(\eta)\left[ - (1+2\phi)d\eta^2 +(1-2\phi)\left( dr^2 +r^2 d\Omega^2 _{(2)} \right) \right] \,,
\ee
but our final results are gauge-independent. We neglect vector and tensor perturbations in the
metric to this order, and spatial derivatives $\partial_i \phi$
of the post-Friedmannian perturbation potential are
assumed to dominate over the time derivative $\partial_t \phi$.

\item The background is assumed to be a spatially flat
FLRW universe dominated by a single dark energy
fluid with energy density $\rho_E$, pressure $P_E$
and equation of state parameter $w \equiv \frac{P_E}{\rho_E} < - \frac{1}{3}$
(until stated explicitly, we do not assume
that $w$ is constant in time or redshift).

\item The spacetime is spherically symmetric, $\phi = \phi (r)$.
This last assumptions recurs in previous
literature \cite{PT, PTT} and is regarded 
here only as a simplification to be relaxed at a later stage, whose 
effects are discussed in \cite{Barrow}.
\end{enumerate}

These assumptions exclude the alternative explanations of the present acceleration of the universe mentioned in the Introduction and the present discussion is to be seen as a test of the $\Lambda$CDM model only. For example, even remaining within the context of General Relativity, if backreaction of inhomogeneities is invoked to explain the cosmic acceleration, the background metric is affected and is no longer given by eq.~(\ref{CNG}), and an effective turnaround radius would have to be introduced through an averaging of the physical variables (which we leave for future work).

\subsection{Hawking-Hayward quasi-local mass and its local and cosmological parts}
The physical, gravitating mass of a non-asymptotically 
flat spacetime has been the subject of much debate
in General Relativity and, after extensive work on
quasi-local energies \cite{Szabados}, the community
seems to have settled on the Hawking-Hayward construct \cite{Hawking, Hayward}.
There is little doubt that
when a region of spacetime of size not entirely negligible
in comparison with the Hubble radius $H^{-1}$ is considered,
a relativistic (as opposed to Newtonian) mass-energy
causing gravitational effects needs to be employed.
The Hawking-Hayward quasi-local mass in defined as
follows \cite{Hawking, Hayward, Hayward2}. Let $S$ be a closed, orientable, 2-dimensional 
surface. Let $\mathcal{R}$ be the induced Ricci scalar on $S$
and consider the outgoing and ingoing congruences of
null geodesics from $S$, which have expansion
scalars $\theta_\pm$ and shear tensors $\sigma_{ab} ^{(\pm)}$, respectively.
Let $\omega^a$ be the projection onto $S$ of the commutator
of the null normal vectors to $S$ (``anholonomicity''),
then the quasi-local mass of $S$ is
\be
M_\text{HH} \equiv \frac{1}{8 \pi} \sqrt{\frac{A}{16 \pi}} \int \mu \left( \mathcal{R} + \theta_+ \theta_- -\frac{1}{2} \sigma_{ab} ^{(+)} \sigma_{(-)} ^{ab} - 2 \omega_a \omega^a \right) \,,
\ee
where $\mu$ is the volume 2-form of $S$ and $A$ is its area \cite{Hawking, Hayward}.

In the geometry (\ref{CNG}), the Hawking-Hayward  mass of a surface $S$ 
corresponding to a sphere in the unperturbed FLRW
background was computed, to first order in the 
perturbations, in Ref.~\cite{VMA}. The result becomes
particularly simple in spherical symmetry in which
the Hawking-Hayward mass reduces \cite{Hayward2} to the better
known Misner-Sharp-Hernandez mass \cite{MisnerSharp, HernandezMisner}.
The result is \cite{VMA}
\be \label{MSH}
M_\text{HH} = ma + \frac{H^2 R^3}{2} \left( 1-\phi \right)
\ee
where $m$ is the Newtonian mass obtained by 
integrating $\phi$ over a sphere, $m= \int \nabla ^2 \phi \, d^3 \vec{x}$,
and is constant. 
The physical mass scale of the perturbation described by $\phi$,
like physical length scales, is obtained
by multiplying by the scale factor (more on this later).
In units in which $G = c = 1$, $2 m$ gives the comoving Schwarzschild 
radius scale of the perturbation, while the physical
scale is $2 m a$. Since $\phi \simeq - m/r \simeq - m a/R$
to first order, where 
\be
R = a r \sqrt{1- 2 \phi} \approx a r (1 - \phi)
\ee
is the areal radius of the perturbed FLRW space in spherical 
symmetry, we can make the approximation
\be \label{approx}
M_\text{HH} \simeq ma + \frac{H^2 R^3}{2}
\ee
to this order. Note that the factor $HR$, the ratio of
the size of a sphere of areal radius $R$ to the Hubble radius $H^{-1}$,
is small for the structures that we will consider, and it appears
to the second power in eq.~(\ref{MSH}). The decomposition (\ref{approx}) splits
the Hawking-Hayward/Misner-Sharp-Hernandez mass into
a ``local'' part $ma$ and a ``cosmological'' part $H^2R^3/2$.
It was shown in \cite{VMA} that this decomposition is
covariant and gauge-invariant to first order, in spite of starting with the
particular gauge (\ref{CNG}).

Eq.~(\ref{approx}) is particularly suited for quantifying the competition between cosmological
expansion and local physics, since the gravitational effects
are due to the physical gravitating masses of the local
structure and of the surrounding cosmology. The criterion
defining the {\it critical} (turnaround) radius for a system
on the verge of breaking down is now that for 
such a system the two contributions to $M_\text{HH}$ are equal,
\be \label{equal}
m a = \frac{H^2 R^3}{2}
\ee
which yields
\be
R_c (t) = \left( \frac{2 m a}{H^2} \right)^{1/3}
\ee
for the critical radius. With this definition there are no
ambiguities in the concept of mass contained in the
sphere of radius $R_c$ (more about this later). Note that eq.~(\ref{equal})
equating the two contributions to $M_\text{HH}$ makes it explicit that
we are in an intermediate regime in which local
gravitational effects are comparable with cosmological ones.
This is the ``gray'' area lying in between the Newtonian
regime $R \ll H^{-1}$ in which cosmological effects are
negligible, and the FLRW regime $R \sim H^{-1}$ in which
relativistic cosmology dominates. 
By using the Hamiltonian constraint
\be
H^2 = \frac{8 \pi}{3} \rho_E
\ee
for a spatially flat FLRW background dominated by dark
energy, one obtains
\be \label{critical_radius}
R_c (t) = \left( \frac{3 m a}{4 \pi \rho_E} \right)^{1/3} \,.
\ee
Let us assume now that the equation of state parameter $w$ is constant; then using the relations valid for the FLRW background
\be \label{scale_factor}
a(t) = a_* t^{\frac{2}{3(1+w)}} \,, \qquad 
\rho_E (t) = \frac{\rho_0}{a^{3(1+w)}} \qquad (w \neq -1) \,,
\ee
where $a_*$ and $\rho_0$ are constants, the critical radius is seen
to vary with the scale factor as
\be
R_c (a) =  \left( \frac{3 m }{4 \pi \rho_0} a^{3w + 4} \right)^{1/3} \equiv R_0 \, a^{\frac{3w+4}{3}}
\ee
(which makes it clear that the ``critical sphere'' is not comoving)
and where the constant $R_0$ is
\be
R_0 \equiv  \left( \frac{3 m}{4 \pi \rho_0} \right)^{1/3} =  \left[ \frac{9 m (1+w)^2}{2 a_* ^{3(1+w)}} \right]^{1/3} \,,
\ee
where the relation
\be
H = \frac{2}{3(1+w)t} = \frac{2 a^{\frac{3(1+w)}{2}}}{3(1+w) a_* ^\frac{3(1+w)}{2}}
\ee
has been used. By showing explicitly the physical local mass $ma$,
one has instead
\be
R_c (a) =  \left( \frac{3 m a}{4 \pi \rho_0} \right)^{1/3} a^{w+1} \,.
\ee
As a function of time, the critical radius is
\be \label{Rct}
R_c (t) = \left( R_0 a_* ^{\frac{3w+4}{3}} \right) t^{\frac{2(3w+4)}{9(w+1}}
\ee
for $w \neq -1$. For $w=-1$, corresponding to a de Sitter background,
it is instead
\be
R_c (t) = R_0 a^{1/3} = R_0 \, a_*^{1/3} \, \mbox{e}^{H_0 t/3} \,.
\ee
The exponent $\frac{2(3w+4)}{9(w+1)}$ in eq.~(\ref{Rct}) is
\begin{itemize}
\item positive if $w < -\frac{4}{3}$ or $w>-1$,
\item negative if $-\frac{4}{3} < w <-1$,
\end{itemize}
and vanishes in a phantom universe with $w=-4/3$,
in which $R_c$ remains constant (a rather odd occurrence).
We can express $R_c$ as a function of redshift $z$ using
$a_0/a = z+1$, where $a_0$ is the present
value of the scale factor. The result is
\be
R_c (z) = \left[ \frac{3 a_0 ^{3(w+1)}}{4 \pi \rho_0} ma \right]^{1/3} \frac{1}{\left( z+1 \right)^{w+1}} = \frac{R_c (0)}{\left( z+1 \right)^{w+4/3}}
\ee
if we insist on using $ma$ as the local mass, where
\be
R_c (0) = \left( \frac{3 m \, a_0 ^{3w+4}}{4 \pi \rho_0} \right)^{1/3} \,.
\ee

By setting the present value $a_0$ of the scale factor to unity,
as customary, one obtains the equation of state parameter
of dark energy as a function of the redshift and the mass $ma$
\be \label{w_z}
w(z) = -1 + \frac{1}{\ln(z+1)} \left\{ \ln \left[ \left( \frac{3ma}{4 \pi \rho_0} \right)^{1/3} \frac{1}{R_c (z)} \right] \right\} \,,
\ee
which can be used to constrain $w$ if $ma$ and $R_c$ are known.

The Hawking-Hayward mass of a ``critical sphere'' of radius $R_c$ is
\be
M_c \equiv M_\text{HH} (R_c) = ma + \frac{H^2 R_c^3}{2} = 2ma = H^2 R^3 _c = \frac{8 \pi}{3} \rho R^3 _c
\ee
by definition, and it includes both the ``local'' mass
$ma$ due to the perturbation and the contribution of the energy
density of the cosmic fluid. When $m$ is constant, 
$M_\text{HH} (R_c)$ is ``comoving'', while $R_c$ is not.

\subsection{Comparison with the previous definition of $R_c$}
We are now ready to compare our $R_c (t)$ with previous
definitions. Since the ``mass''  $\mathcal{M} (r)$ contained in a sphere
of areal radius $R$ is not clearly defined in previous literature,
care must be taken in this comparison. The authors
of \cite{PTT} obtain 
\be
R_c ^{(1)} = \left( \frac{3 \mathcal{M} (r)}{4 \pi |1+3w| \rho_E} \right)
\ee
where it is argued that $\mathcal{M} (r) = 4 \pi \int _0 ^r r'^2 \rho(r') dr'$
and $\rho$ includes the homogeneous energy density (but not the pressure)
of dark energy and the density of the perturbation.
Based on our previous discussion, this would imply
the correspondence $\mathcal{M} \leftrightarrow 2ma$ and
\be \label{ratio}
\frac{R_c}{R_c ^{(1)}} = \left( \frac{|1+3w|}{2} \right)^{1/3} \,.
\ee
For $w \approx -1$ this ratio is almost unity. Therefore,
our calculation can be regarded as a rigorous justification
of the expression of $R_c$ previously obtained in \cite{PTT}
(with a small correction), plus an important clarification 
of the meaning of ``mass in the sphere of radius $R_c$''.
Since the idea is to constrain $w$, however, it is best not
to assume the value of this quantity and to use eq.~(\ref{critical_radius}) for the value of $R_c$ and
eq.~(\ref{ratio}) for its relation with previous literature.

If the equation of state parameter $w$ is not constant, using
the covariant conservation equation
\be \label{conservation}
\dot{\rho}_E + 3 H (P_E + \rho_E) = 0 \,,
\ee
differentiating the relation $a_0/a = z+1$ to
obtain $dt = - \frac{a_0 dz}{\dot{a} (z+1)^2}$, and integrating eq.~(\ref{conservation}) yield
\be
\int dz \, \frac{w(z) +1}{z+1} = \ln \left[ \left( \frac{3ma}{4 \pi \rho} \right)^{1/3} R_c ^{-1} \right]
\ee
which, for $w=$~constant, reduces to eq.~(\ref{w_z}).

\section{Dynamics of the critical sphere}
Let us consider the ``critical sphere'' of radius $R_c (t)$, which
contains a system on the verge of breaking down under
the influence of the cosmic acceleration. Without assuming $w=$~constant,
the differentiation of eq.~(\ref{critical_radius}) yields, in conjunction with
the covariant conservation equation (\ref{conservation}),
\be \label{rate}
\frac{\dot{R}_c}{R_c} = \left( w+ \frac{4}{3} \right) H \,,
\ee
which quantifies the deviation of the critical sphere from
the comoving motion of the cosmic substratum. 
In an accelerated universe with $w <- 1/3$, the critical
sphere grows slower than the cosmic substratum, while
it grows faster in a decelerated universe in which it is 
easier for large structures to collapse. By further
assuming $w=$~const. and using the scale factor~(\ref{scale_factor}),
eq.~(\ref{rate}) becomes
\be
\frac{\dot{R}_c}{R_c} = \frac{2(w+4/3)}{3(w+1)} \, \frac{1}{t}
\ee
for $w \neq -1$, which tends to zero as $t \rightarrow + \infty$ and integrates to
\be
R_c (t) = R_* t^{\frac{2(w+4/3)}{3(w+1)}} \,.
\ee
In the de Sitter case $w=-1$, instead,
\be
R_c (t) = R_* \exp \left[ \left( w + \frac{4}{3} \right) H_0 t \right]
\ee
where $H_0 \equiv \sqrt{\Lambda/3} =$~const.

The physical volume $V_c$ of the critical sphere, of course, obeys
\be
\frac{\dot{V}_c}{V_c} = 3 \, \frac{\dot{R}_c}{R_c} = \left( 3w+4 \right) H \,.
\ee
The Hawking-Hayward/Misner-Sharp-Hernandez quasi-local
mass contained in the critical sphere is
\be \label{Mc}
M_c = \frac{8 \pi}{3} \rho_E R_c ^3 = H^2 R_c ^3
\ee
and its time derivative is easily found to be
\be \label{Mc_dot}
\dot{M}_c = ( H R_c)^3 = \frac{6ma}{H^{-1}}
\ee
using eqs.~(\ref{conservation}) and (\ref{rate}) and without assuming $w=$~const.
Thus, $R_c$ is the fraction $(\dot{M}_c)^{1/3}$ of the Hubble radius
$H^{-1}$. Taking the ratio of eqs.~(\ref{Mc_dot}) 
and~(\ref{Mc}) yields
again the result that $M_c$ is comoving, $ \dot{M}_c/M_c = H$, which
we already know since it was previously established
that $M_c = 2ma$.

\section{Conclusions}

The mathematical splitting of $M_\text{HH}$ gives nearly the same
numerical result as previous literature for the turnaround radius.
This is reassuring since it inspires confidence in the
use of the formal quantity $M_\text{HH}$ in astrophysics, because
its application reproduces almost exactly results obtained with less formal reasoning based on the motion of test and
gravitating particles and fluids, but puts those
results on a firmer theoretical ground. Thus far (with the exception of Ref.~\cite{VMA}), the Hawking-Hayward mass has been used only in formal contexts and computed only 
for analytical solutions of the Einstein equations, but is should be useful also in more practical and astrophysically relevant situations. 

We have shown that the spherically symmetric perturbed FLRW metric is equivalent to a McVittie spacetime and we have provided an exact analysis of the turnaround radius in the latter case. We note that the McVittie metric is a good approximation to the FLRW metric near the turnaround radius as defined by our quasi-local mass splitting, which sits well inside the inner apparent horizon at late times when one exists. 

Following previous literature, we have considered only 
spherical symmetry in this work. The limitations of this
assumption and the quantitative effects of non-sphericity
have been debated in the literature \cite{Barrow}. A background-free discussion without specifying a FLRW background or requiring spherical symmetry should be possible on the lines of Refs.~\cite{Buchert}-\cite{Boleiko}.   The
splitting of the Hawking-Hayward quasi-local mass in
\cite{VMA} is general, and will be applied to non-spherical 
situations in future work.





\acknowledgments
This work is supported by the Natural Science and Engineering Research Council of Canada and by Bishop's University.


\end{document}